\documentstyle[aps,preprint]{revtex}
\begin{document}
\draft
%
\preprint{\vbox{ \hfill SNUTP 95--082}}
\title{$g_{K N \Lambda}$ and $g_{K N \Sigma}$ from QCD sum rules}
\author{Seungho Choe\thanks{E-mail: schoe@phya.yonsei.ac.kr},
        Myung Ki Cheoun\thanks{E-mail: cheoun@phya.yonsei.ac.kr},
        Su Houng Lee\thanks{E-mail: suhoung@phya.yonsei.ac.kr}}
\address{Department of Physics\\
Yonsei University, Seoul 120--749, Korea}
\maketitle
\begin{abstract}
 $g_{K N \Lambda}$ and $g_{K N \Sigma}$ are calculated using a QCD sum
 rule motivated method used by Reinders, Rubinstein and Yazaki to
extract Hadron couplings to goldstone bosons.
The  SU(3) symmetry breaking effects are taken into account by
including the contributions from  the strange quark mass and assuming
different values for the strange and the up down  quark condensates.
We find $g_{K N \Lambda}/\sqrt{4 \pi} = - 1.96 $ and
 $g_{K N \Sigma}/\sqrt{4 \pi} = 0.33 $
\end{abstract}
\vspace{1cm}
%
\section{Introduction}

Over the years, there has been a continuous interest in the
field of kaon-nuclear physics, which range from trying to understand
simple processes like the kaon nucleon scattering or the photo-kaon
production on a nucleon to the spectroscopy and structure of
hypernuclei\cite{Hukuda}. Compared to the pions, the conservation
of strangeness leads to very different interaction of the Kaons
to the nucleon and the nuclei and can therefore yield many
exotic states in nuclear physics \cite{Hukuda} by either a hadronic
or an electromagnetic process.

To understand both these processes and other phenomena in
kaon-nuclear physics, it is important to know hadronic coupling
strengths involving the kaons. Among them, $g_{KN \Lambda}$ and
 $g_{KN \Sigma}$ are the most relevant coupling constants.

 For the pions,  their hadronic coupling constant  $g_{\pi NN}$ is
determined quite accurately through either the nucleon-nucleon
scattering
or the pion-nucleon scattering experiment.  However, the situation
 for the kaons are not as satisfactory. For example, to theoretically
reproduce the experimental kaon-nucleon scattering cross-section,
one usually calculates the contributions from the one-boson exchanges,
the resonances in the s-channel, such as the $\Lambda$ and $\Sigma$,
and the next to leading two meson exchanges\cite{Buettgen}.
These involves many phenomenologically undetermined coupling
constants so that it seems a formidable task to determine the
coupling constants related to the kaons separately.

As another approach, there have been  many
attempts\cite{Buettgen,ABW,AS1,Thom,Tanabe}
to determine these coupling constants from the kaon photo
production.  For instance Adelseck {\it et al.} \cite{ABW,AS1} tried
to
determine these coupling constants($g_{KN \Lambda}$ and
 $g_{KN \Sigma}$) phenomenologically from the data using a
 least-square
 fit method similar to that of Thom's \cite{Thom} and  deduced
some values. But due to the simultaneous determination of many other
unknown coupling constants, these results turned out to have large
uncertainties, i.e. their extracted values
of {{$g_{KN \Lambda} / \sqrt{4 \pi}$} range  from - 1.29 to - 4.17.
Hence, given the uncertainties and difficulties in extracting the
strength of these couplings from the experiments, it is necessary
to explore theoretical predictions.

In this paper, we  will use QCD sum rules \cite{SVZ79} to extract
these kaon couplings.
QCD sum rule is an attempt to understand hadronic parameters in the
low energy region in terms of QCD perturbation theory and
non-vanishing condensate, which characterizes the non-perturbative
QCD vacuum.   This is possible by looking at the correlation function
 between
either two or three QCD hadronic currents and studying its
dispersion relation,  the {\it Real} part of which is calculated in
QCD using the Operator Product Expansion (OPE)  and the {\it Imaginary}
part is modeled with the phenomenological parameters.   This method has
been applied successfully to  the pseudo-scalar hadron hadron
trilinear couplings by Reinders,
Rubinstein and Yazaki\cite{RRY81}, who obtained interesting formulas
such as,
 $g^2_{\pi NN }/4 \pi \simeq 2^5 \pi^3 f_\pi^2/M_N^2$ and
 $g_{\omega \rho \pi} \simeq (2 /f_\pi)(e/2 \sqrt{2})$,
which are numerically in good agreement with the experiment.
Here, we will try to generalize the method to the  kaons and hypernucleons.
The generalization can be made either  within the SU(3) symmetry or  with
the explicit
SU(3) symmetry breaking effects included.  The former case has already been
given in
Ref.\cite{RRY81} and amounts to calculating the F to D
ratio\cite{Georgi}
in QCD sum rule  within the SU(3) symmetry.
The SU(3) symmetry breaking effects in QCD sum rules are taken into
account by
including the effects of strange quark mass and different values for
the
strange quark condensate $ \langle 0|\bar{s} s|0 \rangle
= 0.8 \langle 0|\bar{u} u  |0\rangle$.  This prescription give
good
description for the $\phi$ and $K^*$ meson masses and their
couplings\cite{RRY}.

In the following sections, we will derive the QCD sum rule result
 for the kaon couplings with explicit
symmetry breaking and compare the numerical estimates with the results
 from
phenomenological fitting analyses\cite{Martin,Antolin,AS2,MBH}
 and that of other QCD inspired model
calculations\cite{SM1,SM2}.
%
\section{ QCD sum rules for $g_{KN \Lambda}$}

We will closely follow the procedures given in Ref.\cite{RRY81}.
Consider the three point function constructed of the two baryon currents
 $\eta_B$, $\eta_{B'}$ and the pseudoscalar meson current $j_5$
(Fig. 1)
\begin{eqnarray}
\label{corr}
A(p,p',q) = \int dx\, dy\, \langle0| T (\eta_{B'}(x)j_5(y)
\overline{\eta}_B(0))|0
\rangle\, e^{i(p'\cdot x - q\cdot y)} ,
\end{eqnarray}
In order to obtain $g_{KN\Lambda}$, we will use the following
extrapolating
 fields for the nucleon and the $\Lambda$.
\begin{eqnarray}
\label{current1}
\eta_N&=&\epsilon_{abc}(u_a^T C\gamma_\mu u_b)\gamma_5
\gamma^\mu d_c  ,
\end{eqnarray}
\begin{eqnarray}
\eta_\Lambda&=&\sqrt{\frac{2}{3}} ~\epsilon_{abc} \left[
(u_a^T C\gamma_\mu s_b)\gamma_5\gamma^\mu d_c -
(d_a^T C\gamma_\mu s_b)\gamma_5\gamma^\mu u_c \right]  ,
\label{etalam}
\end{eqnarray}
where u and d are the up and down quark fields ($a,b$ and $c$ are
color
indices),
 $T$ denotes the transpose in Dirac space, and $C$ is the charge
conjugation
matrix.
 For the $K^-$ we choose the current
\begin{eqnarray}
j_{K^-} = \bar{s}i\gamma_5 u  .
\end{eqnarray}
Assuming a pseudo-vector coupling between the nucleon, the $K$ and
the  $\Lambda$, we
expect the following phenomenological form for Eq.(\ref{corr}).
\begin{eqnarray}
\label{phen1}
\lambda_N \lambda_\Lambda \frac{M_B}{(p^2-M_N^2)(p'^2-M_\Lambda^2)}
(\rlap{/}{q} i\gamma_5) g_{KN\Lambda} \frac{1}{q^2-m_K^2}
\frac{f_K m_K^2}{2 m_q} ,
\label{phe}
\end{eqnarray}
%
where $M_B=\frac{1}{2} (M_N + M_\Lambda)$,
 $\lambda_N$ and $\lambda_\Lambda$ are the couplings of the baryons
to their currents.
 $m_q$ is the average of the quark masses,
 $f_K$ is  the kaon decay constant and $m_K$ the kaon mass.
 $M_N$ and $M_\Lambda$ are the masses of the nucleon and the $\Lambda$
particle respectively.
There are other contributions from excited baryon states that couple
to the baryon extrapolating current.
However, we will only look at the pole structure  $\rlap{/}{q}/q^2$ at
 $ q \rightarrow 0 $ and  make a borel transformation to
both $p^2,p'^2 \rightarrow M^2$.  Then, the contributions from the
excited
baryons will be
exponentially suppressed and consequently neglected in our
approximation\cite{RRY}.

As for the OPE side, the perturbative part does not contribute to the
 $\rlap{/}{q}/q^2$ structure.  This is so because the
dimension of Eq.(\ref{corr}) is 4 and $\rlap{/}{q}$ takes away one
dimension such that  only the odd dimensional operators can
contribute.   The lowest dimensional operator with dimension 3 is
the quark
condensate term with higher dimensional operators having the form of
a  quark
condensate with certain number of gluon operators in between.
In fact, for the case of the pions, taking into account only the
leading quark condensate $\langle 0|\bar{q} q |0\rangle$
in the OPE, gives an excellent value for
 $g_{\pi NN}$\cite{RRY}.   Motivated by this result, we will work
out similar
leading
quark condensate contribution as in the pion, which in this case
includes the
contribution from $\langle 0|\bar{s} s|0\rangle$, and further work
out the  additional SU(3) breaking terms up to ${\cal O}(m_s^2)$ and
dimension 7.

 First, we will include the contribution proportional to the $m_s^2$
 in
the Wilson coefficient of the quark condensate.  This will have the
 following
 form,
\begin{eqnarray}
A (p, p', q) = C_u \langle0| \bar{u}u |0\rangle +
               C_d \langle0| \bar{d}d |0\rangle +
               C_s \langle0| \bar{s}s |0\rangle + \cdots\  .
\end{eqnarray}
One can easily show that $C_d = 0$ and
\begin{eqnarray}
C_u&=&-~\sqrt{2\over3} \frac{11p^2}{24\pi^2}
\frac{\rlap{/}{q}}{q^2}
(i\gamma_5)
\ln \frac{\Lambda^2}{-p^2} ,
\label{cu}
\end{eqnarray}
\begin{eqnarray}
C_s&=&-~\sqrt{2\over3} \left(\frac{11p'^2}{24\pi^2} +
\frac{11m_s^2}{48\pi^2} \right)
\frac{\rlap{/}{q}}{q^2} (i\gamma_5)
\ln \frac{\Lambda^2}{-p'^2} ,
\label{cs}
\end{eqnarray}
where $\Lambda$ is the cut-off from the loop integration.
Taking the limit $p'^2 \rightarrow p^2$ and assuming
 $\langle0|\bar{u}u|0\rangle = \langle0|\bar{d}d|0\rangle=
\langle0|\bar{q}q|0\rangle$
 and $\langle0|\bar{s}s|0\rangle = 0.8 ~\langle0|\bar{q}q|0\rangle$
we obtain,
\begin{eqnarray}
\label{twoq}
C_u\langle0|\bar{u}u|0\rangle +C_s\langle0|\bar{s}s|0\rangle =
-~\sqrt{2\over3} \left(\frac{33p^2}{40\pi^2} + \frac{11m_s^2}{60\pi^2}
\right)
 \frac{\rlap{/}{q}}{q^2}
(i\gamma_5) \ln \frac{\Lambda^2}{-p^2} \langle0|\bar{q}q |0\rangle ,
\label{ope}
\end{eqnarray}
Next, we consider the lowest order terms that are
proportional to the strange quark mass $m_s$, namely the dimension 7
operators
 of the type
 $\sim m_s \langle0|\bar{s} s|0\rangle \langle0|\bar{q} q|0\rangle$.
The largest contribution among them comes from the tree graph of
Fig.2,
which gives the following contribution.
\begin{eqnarray}
\label{fourq}
+~\sqrt{\frac{2}{3}} \frac{m_s}{3} \frac{\rlap{/}{q}}{q^2}
(i\gamma_5) \frac{1}{p^2} \langle0|\bar{s} s|0\rangle
\langle0|\bar{q} q|0\rangle .
\end{eqnarray}
After the borel transformation
the typical ratio between the contribution from Eq.(\ref{fourq}) to
Eq.(\ref{twoq}) is $4 \pi^2 \cdot m_s \langle 0|\bar{q} q |0
\rangle /m_N^4$
 for the relevant borel mass range $M^2 \sim m_N^2$.
The factor of $4 \pi^2$ originates from the fact that Eq.(\ref{twoq})
comes from a loop graph  whereas Eq.(\ref{fourq}) does not.
Despite of this loop factors,
the additional condensate effect suppresses the overall ratio
to less than 5\%.   Other dimensional 7 operators come
 from graphs which contain at least one loop and then the
ratio to Eq.(9) becomes
even smaller and can be neglected.

Using Eq.(\ref{twoq}) and Eq.(\ref{fourq}) for the OPE and
Eq.(\ref{phe}) for the phenomenological side, the sum rule after
borel transformation to $p^2=p'^2$ becomes,
\begin{eqnarray}
\lambda_N \lambda_\Lambda \frac{M_B}{M_\Lambda^2-M_N^2}
\left(e^{-M_N^2/M^2} - e^{-M_\Lambda^2/M^2}\right)
g_{KN\Lambda} \frac{f_K m_K^2}{2 m_q} =
\hspace{5cm}
\nonumber\\
\hspace{5cm}
- ~\sqrt{2\over 3} \left(\frac{33}{40\pi^2} M^4
+ \frac{11m_s^2}{60\pi^2} M^2
+~\frac{m_s}{3} \langle0|\bar{s} s|0\rangle \right)
\langle0|\bar{q}q|0\rangle .
\label{twosides}
\end{eqnarray}
 For $\lambda_N$ and $\lambda_\Lambda$, we use the
values obtained from the
 following baryon sum rules for the $N$ and the $\Lambda$ \cite{RRY}:
\begin{eqnarray}
M^6+bM^2+\frac{4}{3}a^2=2(2\pi)^4\lambda_N^2e^{-M_N^2/M^2}
\label{lamn}
\end{eqnarray}
\begin{eqnarray}
M^6+\frac{2}{3}am_s(1-3\gamma)M^2+bM^2+\frac{4}{9}a^2(3+4\gamma)=
2(2\pi)^4\lambda_\Lambda^2e^{-M_\Lambda^2/M^2}
\label{lamlam}
\end{eqnarray}
Here, $a \equiv -~(2\pi)^2 \langle0|\bar{q}q|0\rangle \simeq
0.5 $~GeV$^3$,
 $b \equiv \pi^2 \langle0|(\alpha_s/ \pi) G^2|0\rangle \simeq
0.17$~GeV$^4$,
and $\gamma \equiv \langle0|\bar{s}s|0\rangle / \langle0|
\bar{q}q|0\rangle
-1 \simeq -~0.2$.
We take the strange quark mass $m_s$ = 150 MeV.\\

It should be noted from
Eqs. (\ref{lamn}) and (\ref{lamlam}) that we can not determine
the sign
of $\lambda_N$ and $\lambda_\Lambda$.  Consequently, we can only
determine
the absolute value of $g_{KN\Lambda}$ from our sum rules.
The sum rule in Eq.(\ref{twosides}) should be used for the relevant
borel mass
 $M \simeq M_B = \frac{1}{2} (M_N + M_\Lambda)$.
 Using this we obtain,
\begin{eqnarray}
|g_{KN\Lambda} / \sqrt{4\pi}| \simeq  1.96  ,
\end{eqnarray}
A  more detailed Borel analysis of
Eq.(\ref{twosides}) gives a similar result with $\pm 30$\%
uncertainty.
The uncertainty quoted here comes from neglecting the continuum
contribution in the phenomenological side.
\section{ QCD sum rules for $g_{KN \Sigma}$}
The current of $\Sigma^\circ$ is defined by \cite{Chiu,Jin}
\begin{eqnarray}
\eta_{\Sigma^\circ}
&=& \frac{1}{\sqrt{2}} ~\epsilon_{abc} \left[
(u_a^T C\gamma_\mu d_b)\gamma_5\gamma^\mu s_c +
(d_a^T C\gamma_\mu u_b)\gamma_5\gamma^\mu s_c \right] ,
\nonumber\\*
&=& \sqrt{2} ~~\epsilon_{abc} \left[
(u_a^T C\gamma_\mu s_b)\gamma_5\gamma^\mu d_c +
(d_a^T C\gamma_\mu s_b)\gamma_5\gamma^\mu u_c \right] .
\label{sig0}
\end{eqnarray}
The second form is more useful in our calculation.
Then, within the same approximation as before, the
OPE side looks as follows:
\begin{eqnarray}
C_u\langle0|\bar{u}u|0\rangle +C_s\langle0|\bar{s}s|0\rangle =
+~\sqrt{2} \left(\frac{3p^2}{40\pi^2} + \frac{m_s^2}{60\pi^2}\right)
\frac{\rlap{/}{q}}{q^2}
(i\gamma_5) \ln \frac{\Lambda^2}{-p^2} \langle0|\bar{q}q|0\rangle .
\label{cucssig0}
\end{eqnarray}
In this case there is no term like $\sim m_s \langle0|\bar{s} s|0
\rangle
\langle0|\bar{q} q|0\rangle$.  This is so because the contribution
of this form coming from the first term in Eq.(\ref{sig0})
cancels that
coming from
the second term.
As can be seen from comparing Eq. (\ref{ope}) to Eq.(\ref{cucssig0}),
the
ratio between the leading term and the correction proportional to
 $m_s^2$ are the same for both cases.

Using a similar form in the phenomenological side as in
Eq.(\ref{phe}),  the sum rule looks as follows.\\
\begin{eqnarray}
\lambda_N \lambda_\Sigma \frac{M_B}{M_\Sigma^2-M_N^2}
\left(e^{-M_N^2/M^2} - e^{-M_\Sigma^2/M^2}\right)
g_{KN\Sigma} \frac{f_K m_K^2}{2 m_q} =
\hspace{5cm}
\nonumber\\*
\hspace{5cm}
+ ~\sqrt{2} \left(\frac{3}{40\pi^2} M^4
+ \frac{m_s^2}{60\pi^2} M^2 \right)
\langle0|\bar{q}q|0\rangle .
\label{twosides2}
\end{eqnarray}
Again for  $\lambda_{\Sigma^\circ}$, we take the value
 from the following sum rule for the $\Sigma$\cite{RRY}:
\begin{eqnarray}
M^6-2am_s(1+\gamma)M^2+bM^2+\frac{4}{3}a^2=2(2\pi)^4 \lambda_\Sigma^2
e^{-M_\Sigma^2/M^2} .
\label{lamsigma0}
\end{eqnarray}

Within the same approximation as before, we
take $M \simeq M_B = \frac{1}{2} (M_N + M_{\Sigma^\circ})$ in
Eq.(\ref{twosides2}).
 This
gives the following value for the coupling.
\begin{eqnarray}
|g_{K^-N\Sigma^\circ} / \sqrt{4\pi}| \simeq  0.33 .
\end{eqnarray}

In our approximation, we have SU(2) symmetry; i.e.
we neglected the up and down quark masses, and assumed
 $\langle0|\bar{u}u|0\rangle = \langle0|\bar{d}d|0\rangle$.
Consequently, we can obtain $g_{KN\Sigma}$
using $\eta_{\Sigma^+}$ and $j_{\bar{K}^\circ}$, where
\begin{eqnarray}
\eta_{\Sigma^+}=\epsilon_{abc}(u_a^T C\gamma_\mu u_b)\gamma_5
\gamma^\mu s_c ,
\end{eqnarray}
\begin{eqnarray}
j_{\bar{K}^\circ} = \bar{s}i\gamma_5 d  .
\end{eqnarray}
In this case,  $C_u$=0 and
\begin{eqnarray}
C_d &=& \frac{p^2}{12\pi^2}
\frac{\rlap{/}{q}}{q^2} (i\gamma_5) \ln \frac{\Lambda^2}{-p^2} ,
\label{cd}
\end{eqnarray}
\begin{eqnarray}
C_s &=& \left(\frac{p'^2}{12\pi^2} + \frac{m_s^2}{24\pi^2}\right)
\frac{\rlap{/}{q}}{q^2}
(i\gamma_5) \ln \frac{\Lambda^2}{-p'^2} .
\label{cs+}
\end{eqnarray}
Then the final expression in the OPE side is
\begin{eqnarray}
C_d\langle0|\bar{d}d|0\rangle +C_s\langle0|\bar{s}s|0\rangle =
+~\left(\frac{3p^2}{20\pi^2} + \frac{m_s^2}{30\pi^2}\right)
\frac{\rlap{/}{q}}{q^2}
(i\gamma_5) \ln \frac{\Lambda^2}{-p^2} \langle0|\bar{q}q|0\rangle .
\label{cdcssig0}
\end{eqnarray}
Again, there is no term  proportional to $\sim m_s
\langle0|\bar{s} s|0
\rangle
\langle0|\bar{q} q|0\rangle$.
Neglecting the difference between $M_{\Sigma^\circ}$ and
 $M_{\Sigma^+}$
in the phenomenological side,
and comparing Eq. (\ref{cdcssig0}) with Eq. (\ref{cucssig0})
we obtain the well known relation from isospin symmetry,
\begin{eqnarray}
g_{K^-N\Sigma^\circ} = \frac{1}{\sqrt{2}} g_{\bar{K}^\circ
N\Sigma^+} .
\end{eqnarray}
Because the contribution of each coefficient is the same, we
obtain this relation
despite of taking $\langle0|\bar{s}s|0\rangle =
0.8 ~\langle0|\bar{q}q|0\rangle$ and including the strange quark mass
correction.  This reflects the SU(2) symmetry within our
approach.
(see Eqs. (\ref{cu}), (\ref{cs}) and Eqs. (\ref{cd}), (\ref{cs+})).
%
\section {Discussion}
The SU(3) symmetry , using de Swart's convention , predicts
\begin{eqnarray}
g_{KN\Lambda} &=& - ~\frac{1}{\sqrt{3}} (3 - 2\alpha_D) g_{\pi NN}
\nonumber \\*
g_{KN\Sigma} &=& + ~(2\alpha_D -1) g_{\pi NN}
\label{swart}
\end{eqnarray}
where  $\alpha_D$ is the fraction of the D type coupling,
 $\alpha_D = \frac{D}{D+F}$.
Using the expression of $g_{\pi NN}$ in Ref. \cite{RRY81}
and comparing the OPE sides only,
we obtain $\alpha_D$ = 7/12 in the SU(3) symmetric limit.   This
limit is denoted by QSR I in table I.

Our case(denoted by QSR II in table I)  does not satisfy
Eq. (\ref{swart})
because of the additional SU(3) symmetry breaking factors in the OPE
and  in the phenomenological side.
 Using the convention by
de Swart\footnote{In fact, there is another convention
\cite{AS1,Nagels}.}
 we get
\begin{eqnarray}
g_{KN\Lambda} / \sqrt{4\pi} &=& -~1.96 ,
\nonumber\\*
g_{KN\Sigma} / \sqrt{4\pi} &=& +~0.33  .
\end{eqnarray}

Comparing QSR I and QSR II, we note that the SU(3) symmetry
breaking effect for the couplings are in the order of 25
 $\sim$  30 \%.   This order is similar to the SU(3) symmetry
breaking effects observed in the vector meson masses or the
square of the couplings  to the electro-magnetic current.

In Ref. \cite{AS1} the ranges for the coupling constants are given by
 fitting  $g_{\pi NN}$ and $\alpha_D$ to experimental data and
allowing for SU(3) symmetry breaking at the 20 \% level.
This gives the
following range:
\begin{eqnarray}
g_{KN\Lambda} / \sqrt{4\pi} &=& -~4.4 ~to ~-~3.0 ,
\nonumber\\*
g_{KN\Sigma} / \sqrt{4\pi} &=& +~0.9 ~to ~+~1.3  .
\label{range}
\end{eqnarray}

Other experimentally extracted values, which are
summarized as I, II and III in table I,
lie within the limits  above, except for
the case denoted by IV.

Comparing these limits with our QSR calculations, we observe that
our values fall short of the experimental limits,
although it is closer than the predictions of the Skyrme model.
However, it should be noted that
the present experimental extraction of the couplings
involve a simultaneous determinations of many other unknown
parameters  and a model dependent subprocesses.   Therefore
it is necessary
to investigate the problem further both theoretically and
experimentally.
%
\acknowledgements

The work of SHL and SC were supported in part by the Basic Science
Research
Institute Program, Ministry of Education of Korea, no.
BSRI-95-2425 and by KOSEF
through CTP at Seoul National University.
The work of MKC was supported by KOSEF.
\newpage
\begin{table}

Table I. Coupling constants, $g_{KN\Lambda}$ and $g_{KN\Sigma}$.
Set I and II are the results from an analyses of the kaon~-~nucleon
scattering. Set III is the result of Adelseck and Saghai from the
analysis of the photo~-~kaon scattering and set IV is the result of
Mart {\it et al.} from the  analysis of the charged $\Sigma$
photoproduction.
SM I and II are the Skyrme Model predictions.
QSR I is a QCD sum rule prediction using $\alpha_D$ = 7/12
in the SU(3) symmetric limit. QSR II is our result including
the SU(3) symmetry breaking effects.
\vspace{0.5cm}
\begin{tabular}{c c c c c c c c c }
Coupling Constants & I \cite{Martin} & II \cite{Antolin} &
III \cite{AS2} &  IV \cite{MBH} &
SM I\cite{SM1} & SM II\cite{SM2} & ~~QSR I & ~~QSR II
\\ \hline
 $g_{KN\Lambda}/\sqrt{4\pi}$ & 3.73$^\dagger$ & 3.53$^\dagger$ &
--~4.17 $\pm$ 0.75 & 0.510 & --~2.17$^\ddagger$ & --~0.67$^{\S} $
& --~2.76 & --~1.96 \\
 $g_{KN\Sigma}/\sqrt{4\pi}$ & 1.82$^\dagger$ & 1.53$^\dagger$ &
 ~~1.18 $\pm$ 0.66 & 0.130 &~~~0.76$^\ddagger$ & ~~0.24$^{\S} $
 & ~~~0.44 & ~~0.33 \\
\end{tabular}
\vspace{0.5cm}
 $^\dagger$ Sign undetermined. \\
 $^\ddagger$ With $f_\pi$~=~54 MeV, e~=~4.84 which give the
experimental
values of N and $\Delta$ masses.\\
 $^{\S}$ With the empirical $f_\pi$ =~93.0 MeV,
and e~=~4.82 which gives
a $\Delta$ - N mass difference.
\end{table}

\vfill
%

%
\newpage
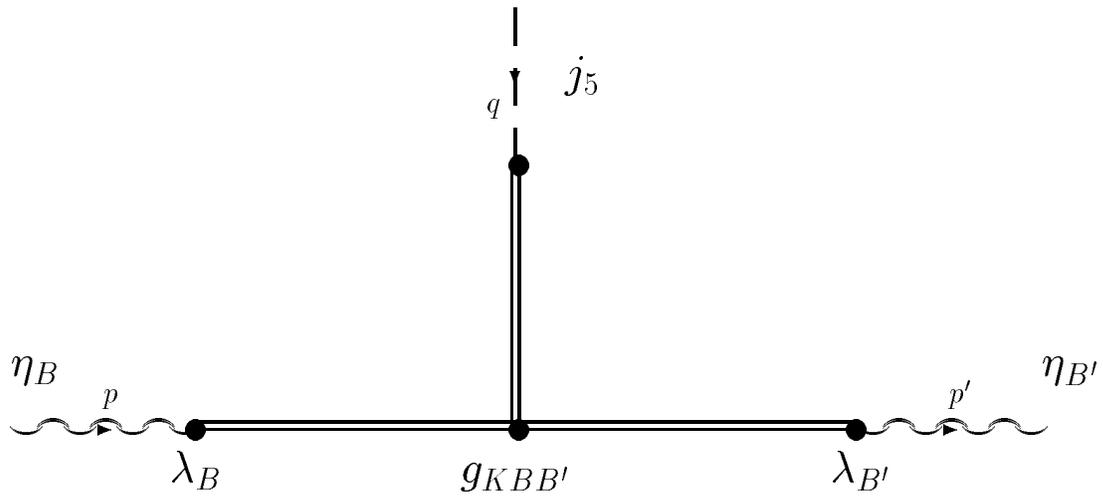
\begin{figure}
\caption{The three point function. $\eta_B$, $\eta_{B'}$ are
the baryon
currents and $j_5$ is the pseudoscalar current. $\lambda_B$ and
 $\lambda_{B'}$ are the couplings of the baryons to the currents, and
 $g_{KBB'}$ is the three point coupling.}
\end{figure}
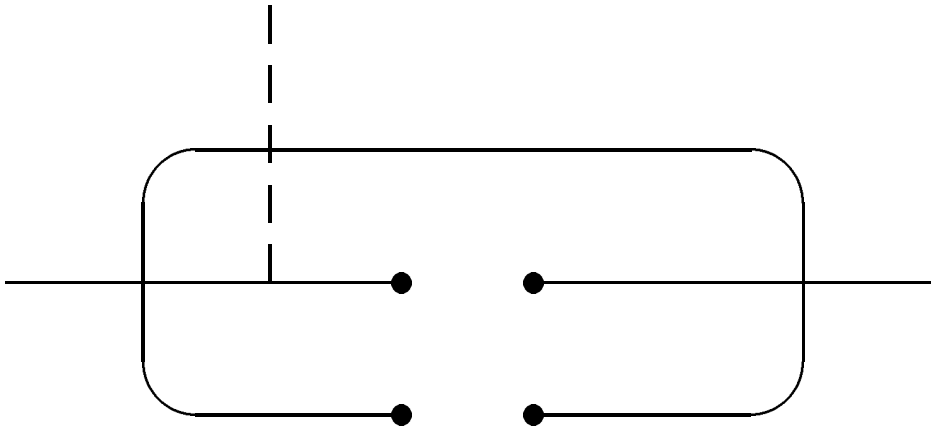
\begin{figure}
\caption{Contribution from the strange quark mass and quark
condensates. Solid lines are the baryon currents and dashed line is
the meson current.}
\end{figure}
%
\end{document}